\documentclass[a4paper,10pt,twocolumn,twoside,aps,prb,superscriptaddress]{revtex4-2}
\usepackage[utf8]{inputenc}
\usepackage{physics}
\usepackage{float}
\usepackage{graphicx}
\usepackage{enumitem}
\usepackage{soul} 
\usepackage[caption=false,subrefformat=parens,labelformat=parens,listofformat=parens]{subfig}
\usepackage{multirow}
\usepackage[utf8]{inputenc}
\usepackage{amsmath}
\usepackage[svgnames]{xcolor}
\usepackage{ulem}
\usepackage[caption=false,subrefformat=parens,listofformat=parens]{subfig}

\begin{document}

\title{Evidence for charge mediated coupling in Fe-Ga/PMN-PT composite multiferroic}
\author{M. Tortarolo$^*$}
\affiliation{Instituto de Nanociencia y Nanotecnolog\'{\i}a (CNEA-CONICET), Nodo Constituyentes, Av. Gral. Paz 1499, B1650KNA San Mart\'{\i}n, Prov. Buenos Aires, Argentina}
\affiliation{Laboratorio Argentino de Haces de Neutrones (CNEA), Centro At\'{o}mico Constituyentes, Av. Gral. Paz 1499, B1650KMA San Mart\'{\i}n, Prov. Buenos Aires, Argentina}
\email{Correspondingauthor: tortarol@tandar.cnea.gov.ar}

\author{D. Goijman$^\dag$} 
\affiliation{Instituto de Nanociencia y Nanotecnolog\'{i}a (CNEA-CONICET), Nodo Bariloche. Centro At\'{o}mico Bariloche, R8402AGP San Carlos de Bariloche, Argentina.}
\affiliation{Laboratorio de Resonancias Magn\'{e}ticas, Comisi\'{o}n Nacional de Energía At\'{o}mica. Centro At\'{o}mico Bariloche, R8402AGP San Carlos de Bariloche, Argentina.}
\thanks{These authors contributed equally to this work}

\author{M. A. Barral$^\dag$} 
\affiliation{Instituto de Nanociencia y Nanotecnolog\'{\i}a (CNEA-CONICET), Nodo Constituyentes, Av. Gral. Paz 1499, B1650KNA San Mart\'{\i}n, Prov. Buenos Aires, Argentina}
\affiliation{Departamento de F\'{i}sica de la Materia Condensada (GIyA-CNEA), Av. Gral. Paz 1499, B1650KNA San Mart\'{\i}n, Prov. Buenos Aires, Argentina}
\thanks{These authors contributed equally to this work}

\author{S. Di Napoli$^\dag$} 
\affiliation{Instituto de Nanociencia y Nanotecnolog\'{\i}a (CNEA-CONICET), Nodo Constituyentes, Av. Gral. Paz 1499, B1650KNA San Mart\'{\i}n, Prov. Buenos Aires, Argentina}
\affiliation{Departamento de F\'{i}sica de la Materia Condensada (GIyA-CNEA), Av. Gral. Paz 1499, B1650KNA San Mart\'{\i}n, Prov. Buenos Aires, Argentina}
\thanks{These authors contributed equally to this work}

\author{A. A. Pérez Martínez} 
\affiliation{Instituto de Nanociencia y Nanotecnolog\'{i}a (CNEA-CONICET), Nodo Bariloche. Centro At\'{o}mico Bariloche, R8402AGP San Carlos de Bariloche, Argentina.}
\affiliation{Laboratorio de Resonancias Magn\'{e}ticas, Comisi\'{o}n Nacional de Energía At\'{o}mica. Centro At\'{o}mico Bariloche, R8402AGP San Carlos de Bariloche, Argentina.}
\affiliation{Universidad Nacional de Cuyo, Instituto Balseiro. Centro At\'{o}mico Bariloche, R8402AGP San Carlos de Bariloche, Argentina.}

\author{G. Ramírez} 
\affiliation{Instituto de Nanociencia y Nanotecnolog\'{\i}a (CNEA-CONICET), Nodo Constituyentes, Av. Gral. Paz 1499, B1650KNA San Mart\'{\i}n, Prov. Buenos Aires, Argentina}
\affiliation{Universidad Argentina de la Empresa (UADE), Instituto de Tecnolog\'{i}a (INTEC), (C1073) Lima 757, Ciudad Autónoma de Buenos Aires, Argentina.}

\author{A. Sarmiento} 
\affiliation{Instituto de Nanociencia y Nanotecnolog\'{i}a (CNEA-CONICET), Nodo Bariloche. Centro At\'{o}mico Bariloche, R8402AGP San Carlos de Bariloche, Argentina.}
\affiliation{Laboratorio de Resonancias Magn\'{e}ticas, Comisi\'{o}n Nacional de Energía At\'{o}mica. Centro At\'{o}mico Bariloche, R8402AGP San Carlos de Bariloche, Argentina.}

\author{J. Gómez} 
\affiliation{Instituto de Nanociencia y Nanotecnolog\'{i}a (CNEA-CONICET), Nodo Bariloche. Centro At\'{o}mico Bariloche, R8402AGP San Carlos de Bariloche, Argentina.}
\affiliation{Laboratorio de Resonancias Magn\'{e}ticas, Comisi\'{o}n Nacional de Energía At\'{o}mica. Centro At\'{o}mico Bariloche, R8402AGP San Carlos de Bariloche, Argentina.}

\author{A. Zakharova}
\affiliation{Swiss Light Source, Paul Scherrer Institut, Forschungstrasse 111 5232 Villigen PSI, Switzerland}

\author{S. E. Bayram}
\affiliation{Swiss Light Source, Paul Scherrer Institut, Forschungstrasse 111 5232 Villigen PSI, Switzerland}

\author{F. Stramaglia}
\affiliation{Swiss Light Source, Paul Scherrer Institut, Forschungstrasse 111 5232 Villigen PSI, Switzerland}

\author{C. A. F. Vaz}
\affiliation{Swiss Light Source, Paul Scherrer Institut, Forschungstrasse 111 5232 Villigen PSI, Switzerland}

\author{ J. Milano}
\affiliation{Instituto de Nanociencia y Nanotecnolog\'{i}a (CNEA-CONICET), Nodo Bariloche. Centro At\'{o}mico Bariloche, R8402AGP San Carlos de Bariloche, Argentina.}
\affiliation{Laboratorio de Resonancias Magn\'{e}ticas, Comisi\'{o}n Nacional de Energía At\'{o}mica. Centro At\'{o}mico Bariloche, R8402AGP San Carlos de Bariloche, Argentina.}
\affiliation{Universidad Nacional de Cuyo, Instituto Balseiro. Centro At\'{o}mico Bariloche, R8402AGP San Carlos de Bariloche, Argentina.}

\author{ C. Piamonteze}
\affiliation{Swiss Light Source, Paul Scherrer Institut, Forschungstrasse 111 5232 Villigen PSI, Switzerland}

\date{\today}

\begin{abstract}
We present an experimental study  of the magnetoelectric coupling (MEC) in the Fe-Ga/PMN-PT thin film multiferroic composite by means of x-ray magnetic circular dichroism (XMCD) and ferromagnetic resonance (FMR). Our measurements show evidence for a charge mediated coupling mechanism, suggested by the asymmetric magnetic remanence (M$_{rem}$) behaviour under opposite electric fields ($\pm$ E) and the asymmetric resonance field (H$_r$) in the FMR measurements. Also, the FMR measurements reveal a perpendicular magnetic anisotropy (PMA), that can be related to an interface charge effect. and it is tunable by E field. \textit{Ab initio} calculations support the existence of a charge mediated coupling at the Fe-Ga/PMN-PT interface.
\end{abstract}

\maketitle

The interest in controlling the magnetization  by an electric field stems from the possibility of designing power efficient and non-volatile magnetic devices \cite{Ramesh2007, Vaz2010}. These devices include voltage-driven magnetic random access memories \cite{Bibes2008,Hu2010,Liu2011}, voltage-tunable RF/microwave magnetic devices \cite{ Liu2009}, logic circuits \cite{Hu2010_2} and more recently, memristive devices \cite{Shen2021}. In this sense, several artificial multiferroic heterostructures of ferroelectric (FE) and ferromagnetic (FM) layers are being studied due to their magnetoelectric coupling (MEC), either by the strain induced by the FE on the FM layer (magnetoelastic coupling) \cite{Thiele2007,Eerenstein2007,Liu2009,Yang2009,Liu2011,Brandlmaier2011,Wang2010-2011,Wu2011} or by the charge screening at the FE/FM interface (magneto electronic coupling) \cite{Molegraaf2009,Vaz2010-2,Weisheit2007,Maruyama2009,Endo2010} upon applying an electric field.  The electric control of the magnetization at room temperature in such heterostructures has been demonstrated in several systems during the last decade. Combined strain-mediated and charge-mediated magneto-electronic MEC was reported in thin Ni/BTO  \cite{Shu2012}, where a thickness dependent voltage modulation of the magnetic behaviour was observed, namely, a strain mediated ME coupling that dominates at larger film thickness and a charge dominated one that prevails at small thicknesses.
In addition, La$_{0.8}$Sr$_{0.2}$MnO$_3$(4 nm)/PZT showed an hysteretic M-E curve at 100 K due to a charge mediated MEC \cite{Molegraaf2009,Vaz2010-2}, while characteristic strain mediated butterfly like M-E curve was observed in a 50 nm thick La$_{0.7}$Sr$_{0.3}$MnO$_3$/PMN-PT heterostructure \cite{Thiele2005,Heidler2015}. Also, the combined charge-mediated and strain-mediated MEC was reported in NiFe(1nm)/PMN-PT \cite{Nan2014} and in Co/PMN-PT heterostructures \cite{Heidler2016}.\\
 The Fe$_{1-x}$Ga$_x$ rare-earth free alloy, characterized by a high magnetostriction \cite{Clark2003,Atulasimha2011,Meisenheimer2021}, is a natural candidate to realize ME strain coupling to a piezoelectric substrate \cite{Phuoc2017,Zhang2018,Bai2018,Wu2011,Jimenez2019,Jahjah2020,Pradham2024}. These works focused on the strain coupling ME effect dependance on the layer thickness, but the coupling due to the interface charge screening was not discussed. In this work, we address the ME coupling mechanisms in Fe$_{0.83}$Ga$_{0.17}$/Pb[(Mg$_{1/3}$Nb$_{2/3}$)O$_3$]$_{0.68}$-[PbTiO$_3$]$_{0.31}$(011) (Fe-Ga/PMN-PT(011)) multiferroic heterostructures by XMCD and FMR to show the presence of strain as well as interface charge screening ME coupling for thin Fe-Ga layers. These findings are supported by \textit{ab initio} calculations.

\begin{figure}[!t]
	    \includegraphics[width=0.48\textwidth]{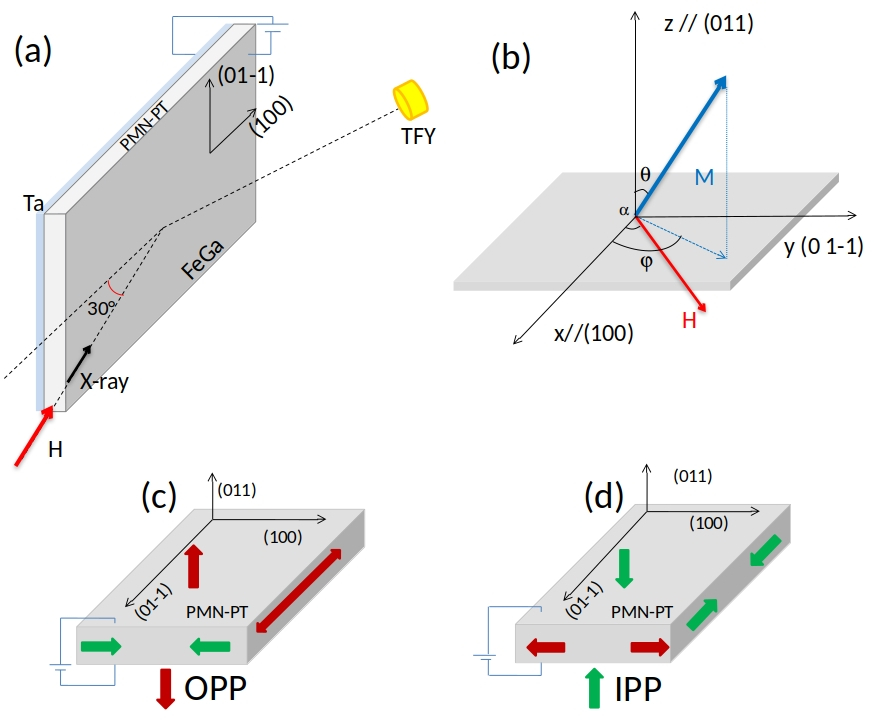}
        \caption{(a) Experimental geometry for the XMCD measurements. (b) Coordinate system relative to the PMN-PT crystal orientation. The Fe-Ga magnetization vector M  and the magnetic field H are described by the polar coordinates ($\varphi$, $\theta$) and $\alpha$, respectively. (c) and (d) show the strain direction in the OPP (c) and IPP (d) configurations.  }
           \label{fig:sketch}
\end{figure}

Polycrystalline Fe-Ga thin films of nominal thicknesses ranging from 2 nm to 8 nm were deposited by magnetron sputtering on 5 mm $\times$ 7 mm $\times$ 0.5 mm PMN-PT (011) crystalline substrates (Atom Optics Co, LTD, Shanghai, China) from a commercial 38 mm diameter target, with a nominal composition of Fe$_{0.75}$Ga$_{0.25}$. Rutherford backscattering spectrometry (RBS) characterization showed that the actual concentration of the films was Fe$_{0.83}$Ga$_{0.17}$ \cite{Ramirez2021}. A 60~nm Ta  layer was deposited on the bottom of the substrate to be used as bottom contact and a 5~nm Pt layer was deposited on top of the magnetic layer to prevent oxidation. Sputtering conditions for the Fe-Ga alloy were: base pressure 
$<$~1.2$\times$10$^{-6}$~Torr, 20~W sputtering power (1.8~W/cm$^2$ power density) and 2~mTorr Ar pressure which gave a sputtering rate of  $\sim$0.15 nm/s.  
XMCD measurements at the Fe L$_{2,3}$ were carried at the ES3 endstation of the SIM beamline \cite{SIM_beamline} at the Swiss Light Source, Paul Scherrer Institut, Switzerland. The XMCD signal, consisting on the absorption intensity difference between opposite x-ray helicities, is an element specific probe of the magnetization along the photon propagation direction \cite{JStor2006}.  Spectra were recorded at room temperature with an incident angle of 60$^{\circ}$  from the surface normal, measuring mostly the projected magnetization along the [100] crystal direction of the PMN-PT (Figure \ref{fig:sketch}), corresponding to the maximal deformation of the crystal and the easy magnetic axis of the magnetic layer, as the characterization of the virgin sample (before applying electric field) confirmed. The external magnetic field was applied parallel to the x-ray beam. Spectra were recorded using  total fluorescence yield (TFY). We estimate that the probing depth of the TFY detection mode at the Fe L$_3$-edge in our
experimental geometry (grazing incidence) is approximately 8~nm (see section \ref{prob-depth}), which
is of the same order as the thickest film measured. Therefore, the contribution of the Fe-Ga layers interfacing
with PMN-PT will contribute significantly more to the total measured signal for the 2~nm film than for the 8~nm film.

FMR measurements were carried out in a Bruker Elexys E500 system working in X-band ($\sim$ 9,5 GHz) at room temperature. The sample is placed in a rectangular resonant cavity.  The derivative of the absorbed power was measured using standard field modulation and lock-in detection techniques with a modulation amplitude of 20 Oe. In both XMCD and FMR experiments, the electric field (in the range $\pm$ 0.3 MV/m) was applied to the bottom electrode with the top electrode connected to ground. 

Three different remanent  FE polarization states are possible in PMN-PT(011): when the FE polarization is poled positively or negatively out of plane (OPP$\pm$) by applying an electric field of $\pm$ 0.3 MV/m, and when the FE polarization is in-plane poled (IPP) at the coercive electric field (E$_{c}$ $\sim$ $\pm$ 0.14 MV/m). In the first case, no lattice parameter change is expected when comparing OPP+ and OPP-, and the Fe-Ga layer experiences identical strain conditions with E field. The switching from OPP to IPP implies structural changes in the PMN-PT \cite{Heidler2015} as depicted in Figure~\ref{fig:sketch}(c) and (d), inducing different strain states on the Fe-Ga layer. Moreover, FE polarization switching at the E$_{c}$ alters the interfacial charge that should be screened by the adjacent Fe-Ga layer by the accumulation or depletion of electrons, which has an impact on the surface anisotropy of the FM layer \cite{Duan2006}. 

\begin{figure}[!h]
    \centering
    \includegraphics[width=1\linewidth]{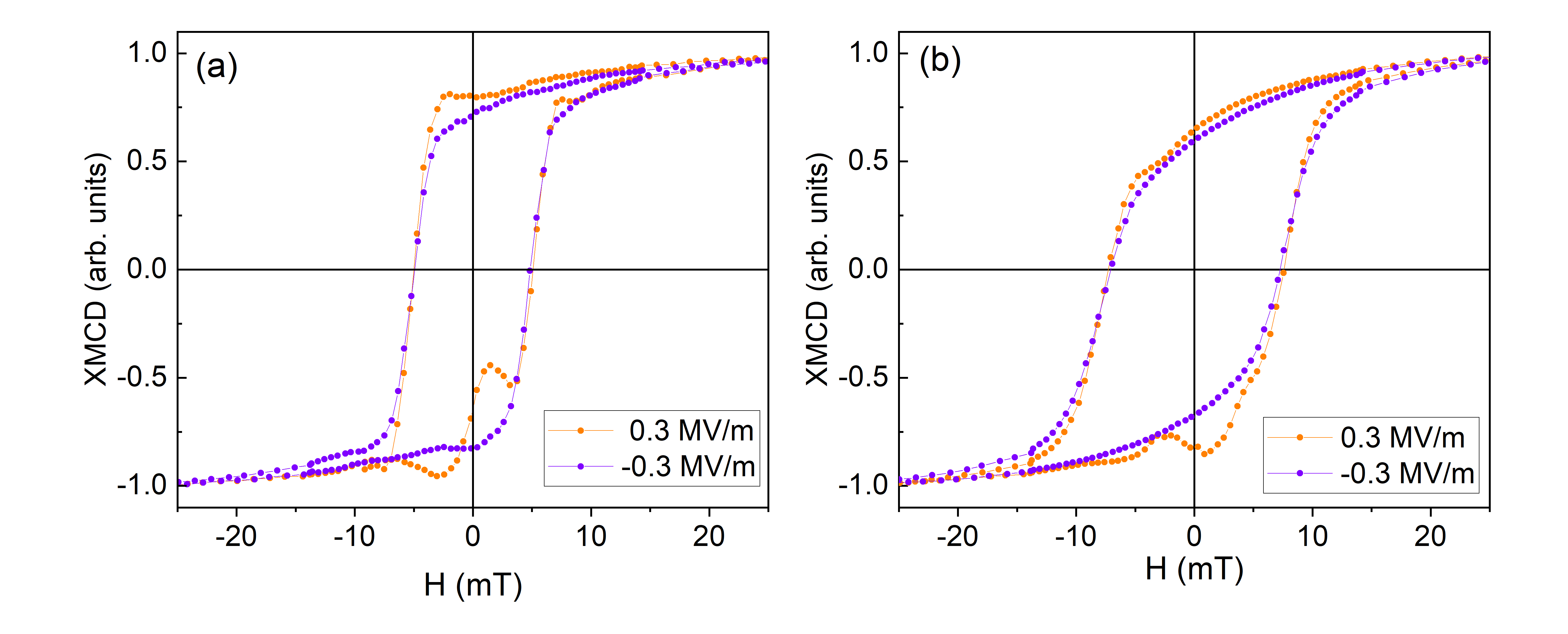}
    \caption{(Color online) XMCD hysteresis loops along the [100] direction for OPP$\pm$ states for the 2 nm  (a) and the 8 nm-thick FeGa film/PMN-PT (b). The data has been smoothed with 5 points average.}
    \label{fig:XMCD}
\end{figure}

Fe XMCD hysteresis loops along the [100] direction measured  for 2 and 8\,nm-thick FeGa/PMN-PT (011) films are shown in Figure \ref{fig:XMCD}. The loops have been normalized to the saturation magnetization ($M_{sat}$). Figures \ref{fig:XMCD} (a) and (b) show the comparison between OPP+ and OPP- for 2\, and 8\,nm-thick films, respectively. As seen in Figure \ref{fig:XMCD}, both films show a difference at the magnetic remanence ($M_{rem}$/$M_{sat}$, the magnetization at zero applied magnetic field) between OPP+ and OPP- states. The differences are better seen in the branch for decreasing magnetic field, as the increasing magnetic field shows noise around zero applied magnetic field. For the 2\,nm-thick  film M$_{rem}$/M$_{sat}$ goes from 0.79(1) to 0.71(2) from OPP+ to OPP-, a decrease of 10\%. For the 8\,nm-thick film $M_{rem}/M_{sat}$ changes from 0.650(4) to 0.599(1), a decrease of 7.8\%. For the thinner film the change is more visible since the loop changes shape, having a more square-like shape as visible by a larger increase from OPP+ compared to OPP- at -2\,mT. Since no structural difference should occur on PMN-PT between OPP+ and OPP- states, the changes observed by XMCD point to an existing charge-mediated magnetoelectric coupling in the FeGa/PMN-PT.

\begin{figure}[!h]
\includegraphics[width=1\textwidth]{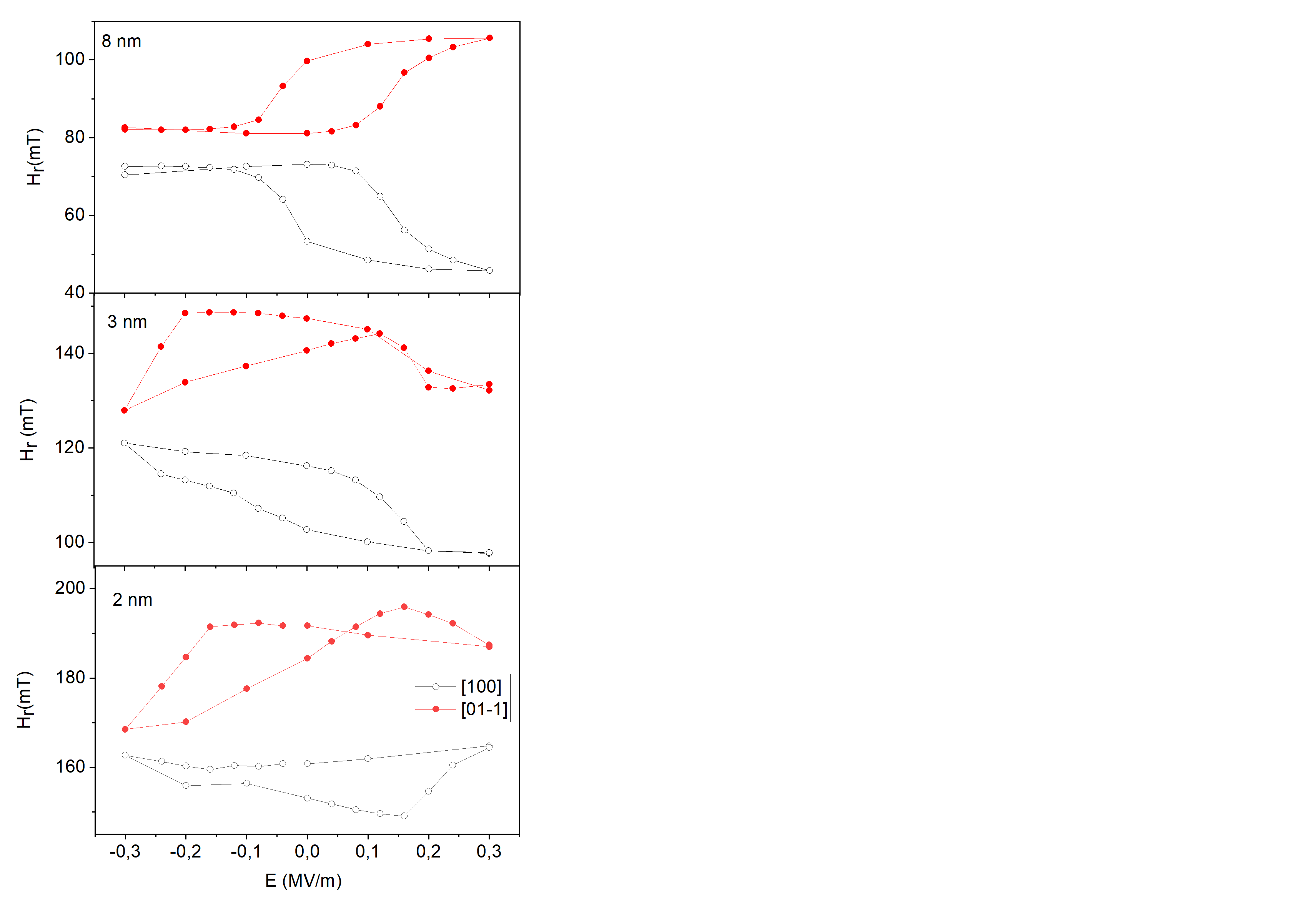}
\caption{$H_r$ as a function of the electric field, measured with the applied magnetic field along the [100] and [01-1] directions (black and red symbols, respectively), for 8~nm (top panel), 3~nm (middle panel), and 2~nm (lower panel).}
\label{fig:Hr}
\end{figure}

To obtain a more quantitative measurement of the change effects in magnetic anisotropy with electric field we have performed FMR measurements on the same films. The magnetic resonance field $H_r$ under an applied electric field $E = \pm 0.3$~MV/m for three different sample thicknesses, measured with the applied magnetic field along [100] and [01-1] directions are shown in Fig.~\ref{fig:Hr}.
The characteristic feature of the charge-mediated ME coupling is an asymmetric $H_r$ behaviour with the $E$ field, similar to that exhibited by the 8~nm- thick sample 
(Fig.~\ref{fig:Hr} (a)), which almost reproduces the ferroelectric loop ~\cite{crystal}. The 3 and 2~nm-thick samples also show asymmetry in H$_r$ vs E loops, while the 2~nm also has superimposed a "butterfly" shape characteristic of strain coupling. Table~\ref{tab:anisotropy_constants} shows the fitting results for the $K_{\rm PMA}$ for different thicknesses at maximal $\pm E$ \ref{FMR fit}. $K_{\rm PMA}$ shows its maximum values for the 3~nm-thick sample. It is important to note that the 2~nm-thick sample presents a different behavior with respect to the other samples. This is particularly observed in the $M_s$ reduction, which could be attributed to charge accumulation and depletion at the Fe-Ga/PMN-PT interface \cite{Avula2018}.
$K_{\rm PMA}$ also shows a variation with $E$: OPP+ has a lower $K_{\rm PMA}$ absolute value than the OPP-, indicating an easier out of plane magnetic anisotropy axis for the OPP- state. 
\\
\begin{table}[h]
\caption{\label{tab:anisotropy_constants}
$K_{\rm PMA}$: Experimental values from the fit (S2)}
\begin{ruledtabular}
\begin{tabular}{lcr}
\textrm{Sample pooling}&
\textrm{$M_s$ (kA/m)}&
\multicolumn{1}{c}{\textrm{$K_{\rm PMA}$ (x10$^{5}$J/m$^3$)}}\\
\colrule
8 nm OPP+ & 1200 & -1.00(0.01)\\
8 nm OPP-& 1200 & -1.50(0.01)\\
3 nm OPP+& 1200 & -4.24(0.04)\\
3 nm OPP-& 1200 & -4.74(0.04)\\
2 nm OPP+& 840 &  -2.47(0.06)\\
2 nm OPP- & 840 & -2.65(0.06)\\
\end{tabular}
\end{ruledtabular}
\end{table}

The thickness variation of the FMR M-E loops suggests therefore, that the charge ME effect is stronger for the thicker FeGa films while for the thinner films, the strain magnetoelastic effect seems stronger. This behaviour is distinct to that found  for FePt/PMN-PT ~\cite{Yang2017} where a clear crossover from magnetoelastic to charge mediated coupling with decreasing film thickness was observed. However, in our case a more complex behavior is expected since, as shown Ref.~\cite{Jimenez2019} the absolute value of the magnetoelastic coupling in Fe-Ga/PMN-PT tends to decrease with increasing film thickness from 6~nm to 20~nm (there is no data for thinner films), and noticeably increases again above 25\,nm. This was attributed to a thickness dependence of the magnetoelastic constant \textit{B}. This could explain why the magnetoelastic contribution for the 8\,nm is smaller than for the 2\,nm film, as observed in figure \ref{fig:Hr}, and no monotonic crossover from charge-mediated to magnetoelastic coupling with increasing film thickness is observed.

\begin{figure*}
\includegraphics[width=1\textwidth]{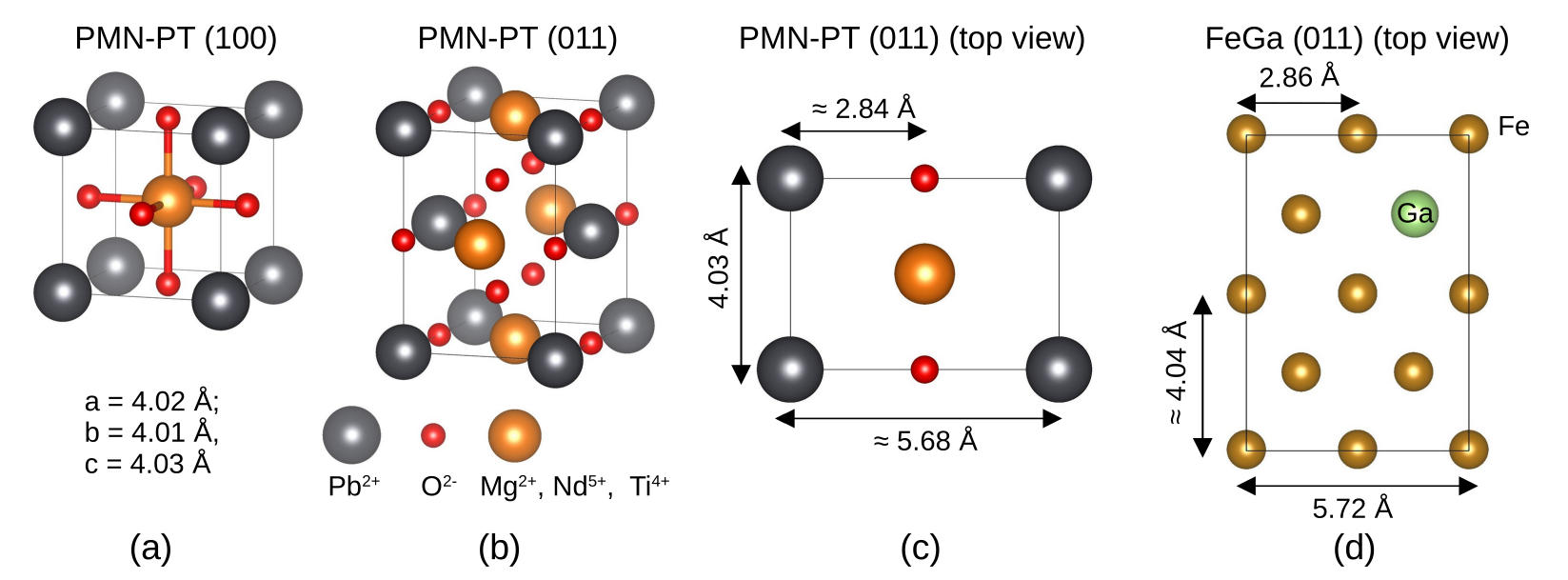}
\caption{(Color online) Schematic representation of (a) PMN-PT(001) and (b) PMN-PT(011) bulk perovskite orientations. (c) Top-view of the (011) surface. (d) Top-view of the modelled (011) Fe$_{0.83}$Ga$_{0.17}$ slab. The figures were generated using the VESTA software package \cite{VESTA}.}
\label{fig:structure}
\end{figure*}

The XMCD results indicate a charge-mediated magnetoelectric coupling  in both film thicknesses even though larger for the 2\,nm film than for the 8\,nm film. For FMR the charge-mediated coupling is quite evident for the 8\,nm film and less so for the 2\,nm film. These differences can be explained by the fact that the probing depth of the two techniques is different. FMR probes the full volume equally while the XMCD from the top surface is stronger than the signal from the interface. As shown in section \ref{prob-depth}, the attenuation length of the x-rays at the Fe resonance is  8.4\,nm. Taking this value into account, we find that the contribution of the layers at the interface to the substrate normalized by the total XMCD signal is almost 6 times larger for the 2\,nm compared to the 8\,nm-thick FeGa film. Since the charge contribution is expected to happen only at the interface, this  explains why XMCD shows a smaller charge contribution for the 8\,nm film while FMR shows a more evident change. The FMR results on the 8\,nm film evidence a lower magnetoelastic response in this film compared to the 2\,nm film. As mentioned above, it has been observed previously for FeGa/PMN-PT that the magnetoelastic effect has a complex dependence on thickness~\cite{Jimenez2019}. The reason for this non-monotonic behavior remains unclear and further investigation is needed.

A complementary perspective of the screening charge effect on the magnetic anisotropy can be obtained from \textit{ab initio} calculations within the density functional theory (DFT). We perform DFT calculations for the fixed strain  given by the OPP$\pm$ poled PMN-PT states, in a model system (see Section~\ref{S-DFT-method} of the SI for details). 
In Figs.~\ref{fig:structure}(a) and (b) we show schematic representations of PMN-PT cubic perovskite in the (100) and (011) orientations, respectively.
When Fe-Ga is grown on top of PMN-PT(011), the substrate imposes its lattice constants and symmetry and, therefore we propose to study a [011]-oriented Fe-Ga film based on the good lattice matching between this film crystallographic orientation and the substrate, as can be seen in Figs.~\ref{fig:structure}(c)-(d). This scenario is consistent with Ref.~\cite{110-textured}, where the sputtered polycrystalline Fe-Ga/PMN-PT(011) films show mostly a (011) textured growth, as revealed by XRD measurements. 
As already mentioned, the deformation of the FE layer is symmetric when switching the ferroelectric polarization (\textbf{P}$_{PMN-PT}$) from OPP+ to OPP- yielding the same strain state, while the screening charges depend on the electric field polarity ~\cite{crystal}. Taking into account that these charges are localized at the interface and will decay exponentially within the Thomas-Fermi screening length, we consider a thin film consisting of only 4 Fe-Ga stacked layers ($\sim$ 6~\AA \, thick) and 12~\AA \, of vacuum  perpendicular to the surface. The surface unit cell consists in a 2$\cross$1 atom array in the [100] and [01-1] directions, respectively (Fig.~\ref{fig:structure}(d)), yielding a total of 26 Fe and 6 Ga atoms in the unit cell, consistent with the experimental Ga concentration in the Fe matrix.

According to Ref.[~\onlinecite{Heidler2015}], switching the FE polarization of the PMN-PT from IPP to OPP leads to a tensile strain of 0.23\% along [01-1] and to a contraction of -0.90\% along the [100] direction. Therefore, we model our experimental Fe$_{0.83}$Ga$_{0.17}$ samples by fixing the lattice constants to those of PMN-PT, that is $a\sim 7.98$\,\AA\, and $b\sim 5.81$\,\AA\,. With 2$\cdot$\textbf{P}$_{PMN-PT}=60\,\mu C/cm^2$~\cite{Heidler2016}, the amount of interface charge doping per surface unit cell can be estimated to be $\pm 1e^-$, which is taken into consideration by repeating our MAE computations varying the total e$^-$ count within the unit cell in the DFT calculations. 

\begin{figure}
\includegraphics[width=1.\columnwidth]{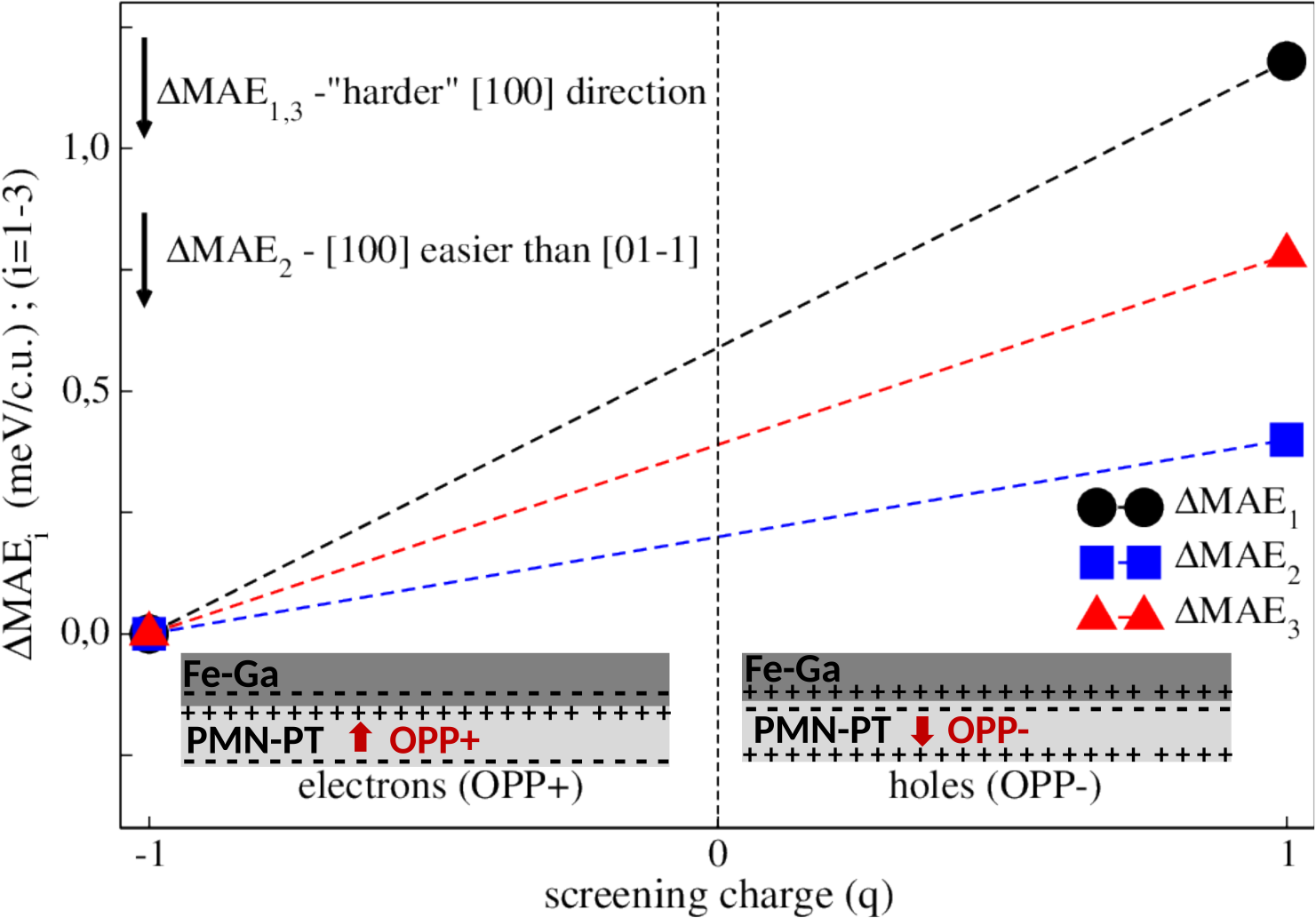}
\caption{Change in MAE with respect to the OPP+ poled one, that is $\Delta MAE_i = MAE_i(OPP\pm)-MAE_i(OPP+)$. Schematic representations of the screening charges for the different polarization orientations are shown in the insets of the figure. The dashed lines are guides to the eyes. }
\label{fig:dft}
\end{figure}

We compute three different MAEs, i.e. MAE$_1$=E$_T$[100]-E$_T$[011], MAE$_2$=E$_T$[100]-E$_T$[01-1] and MAE$_3$=E$_T$[01-1]-E$_T$[011], with E$_T$ the total energy of each configuration, as a function of the screening charge corresponding to the two polarization directions. We recall that when the FE polarization points to the Fe-Ga film (OPP+) the screening charge is negative (electron accumulation), while it is positive (electron depletion) when the polarization points in the opposite direction (OPP-). To facilitate the interpretation of our results and highlight the effect of the screening charge sign, we compare the MAE$_i$ values with respect to the corresponding ones when the screening charge is negative, that is, $\Delta MAE_i = MAE_i(OPP\pm)-MAE_i(OPP+)$. As can be seen in Fig.~\ref{fig:dft}, $\Delta MAE_1$ and $\Delta MAE_2$ are positive, indicating that the [100] axis becomes magnetically harder when switching the polarization from OPP+ to OPP-, in agreement with the experimental behaviour of the $M_r$ as a function of the polarization direction, as shown in Fig.~\ref{fig:XMCD}(c). Furthermore, the fact that $\Delta MAE_1$ and $\Delta MAE_3$ are positive indicates that the perpendicular axis becomes easier when the screening charge is positive, in agreement with FMR experiments.

In summary, we have systematically investigated by XMCD, FMR and \textit{ab initio} calculations the existence of a charge mediated ME coupling in the FeGa/PMN-PT composite multiferroic. XMCD measurements hint the presence of a charge screening effect at the Fe-Ga/PMN-PT interface, which was studied by FMR experiments showing an out of plane anisotropy $K_{\rm PMA}$ that can be controlled with the $E$ field. \textit{Ab initio} calculations confirmed the experimental trends of the anisotropy with the electric field. The existence of an additional charge-mediated ME coupling, besides the already known magnetoelastic one,  in ultra thin FM/FE heterostructures opens the path to enhanced ME coupling, as well as the tailoring of magnetic spin textures through electric field pulses and for the development of efficient low power consumption non volatile magnetoelectric and spintronics devices.

\section*{Acknowledgements}
The authors acknowledge the financial support of the Leading House for the Latin American Region, St. Gallen University under grant SMG 1918. A.Z. acknowledges the financial support by the Swiss National Science Foundation (SNSF) under Project No. 200021\_169467. F. S. is supported by the Swiss National Science Foundation (SNCF) under grant No 200021\_184684. Part of this work was performed at the Surface/Interface: Microscopy (SIM) beamline of the Swiss Light Source (SLS), PSI, Villigen, Switzerland.

\section*{Supplemental Material}
\subsection*{Estimation of probing depth}\label{prob-depth}
We estimate the probing depth of the total fluorescence yield. In~\cite{Regan} the absorption coefficient calculated for Fe metal at the L$_3$-edge resonance is approximately 60\,/$\mu$m, which corresponds to an attenuation length of $\lambda_{res}$=1/60=0.0167\,$\mu$m=16.7\,nm. This value corresponds to the attenuation length for normal incidence. In grazing incidence the attenuation length projected along the normal to the surface is $\lambda_{res} \times cos(\theta)$, where $\theta$ is the angle between the film  normal and the x-ray incidence direction, which in our experiment is 60$^{\circ}$. Therefore, in our measurement conditions the penetration length of the x-rays from the sample surface is $\lambda_{res,graz}=8.3$\,nm.

\subsection*{Calculation method}
\label{S-DFT-method}
First-principles calculations are done within the framework of the density functional theory (DFT) and the projector augmented wave method~\cite{DFT}, as implemented in the Vienna ab initio package (VASP)~\cite{VASP1,VASP2}.  We explicitly treat the Fe (3s, 3p, 3d, 4s) and Ga (3d, 4s, 4p) electrons as valence electrons. The generalized gradient approximation (GGA) in the parametrization of Perdew, Burke, and Ernzerhof (PBE) is employed~\cite{PBE}.
The energy cutoff for the plane-wave basis was set at 420~eV and the interatomic forces is minimized to 0.01 eV/\AA\, for the structural relaxations.

Due to the random disorder of the A2 structure of Fe$_{0.83}$Ga$_{0.17}$, we optimize our supercell obtained from special quasirandom (SQSs)~\cite{sqs-1} 2$\cross$2$\cross$2 BCC structures, with stoichiometry of Fe$_{26}$Ga$_6$,  generated  with the Alloy Theoretic Automated Toolkit~\cite{sqs-2}. Atoms were arranged to approximate the pair-correlation functions of random alloys up to third neighbours distance. From this supercell, we build our (011)-slab, containing 4 stacked layers ($\sim$ 6~\AA \, thick) and 12~\AA \, of vacuum along the perpendicular direction of the surface, in a 2$\cross$1 atom array in the [100] and [01-1] directions, respectively. 

Spin orbit interaction, as implemented in VASP, is included in our calculations and the magnetocrystalline anisotropy energies (MAE) are computed using the force theorem, which allows one to evaluate the energy difference using non-self consistent band energies~\cite{ForceTheorem}. 

\subsection*{Fits to the FMR data}
\label{FMR fit}

The magnetic behaviour of the Fe-Ga film can be described by the following free energy model:

\begin{equation}\label{eq:Free_energy}
\begin{aligned}
U(\theta,\varphi) ={} & \mu_o\textbf{H}.\textbf{M} + \frac{\mu_o}{2} M^2 \cos^2\theta + \\
      & + K_{PMA}\cos^2\theta+K_u\sin^2\theta \cos^2(\varphi-\varphi_u).
\end{aligned}
\end{equation}

 \textbf{H} is the applied magnetic field and \textbf{M} is the magnetization vector. The first term on the right hand is the Zeeman energy. The second one corresponds to the demagnetizing dipolar energy for a thin film. The third therm accounts for a perpendicular anisotropy (K$_{PMA}$), and the fourth one corresponds to the uniaxial magnetic anisotropy $K_u$. This latter term models all the the in-plane magnetic anisotropies that present a uniaxial behavior, such as magnetic anisotropies originated during the growth process and the effective anisotropy as a result of the coexisting crystalline textures within the samples.



\begin{figure}
\includegraphics[width=0.5\textwidth]{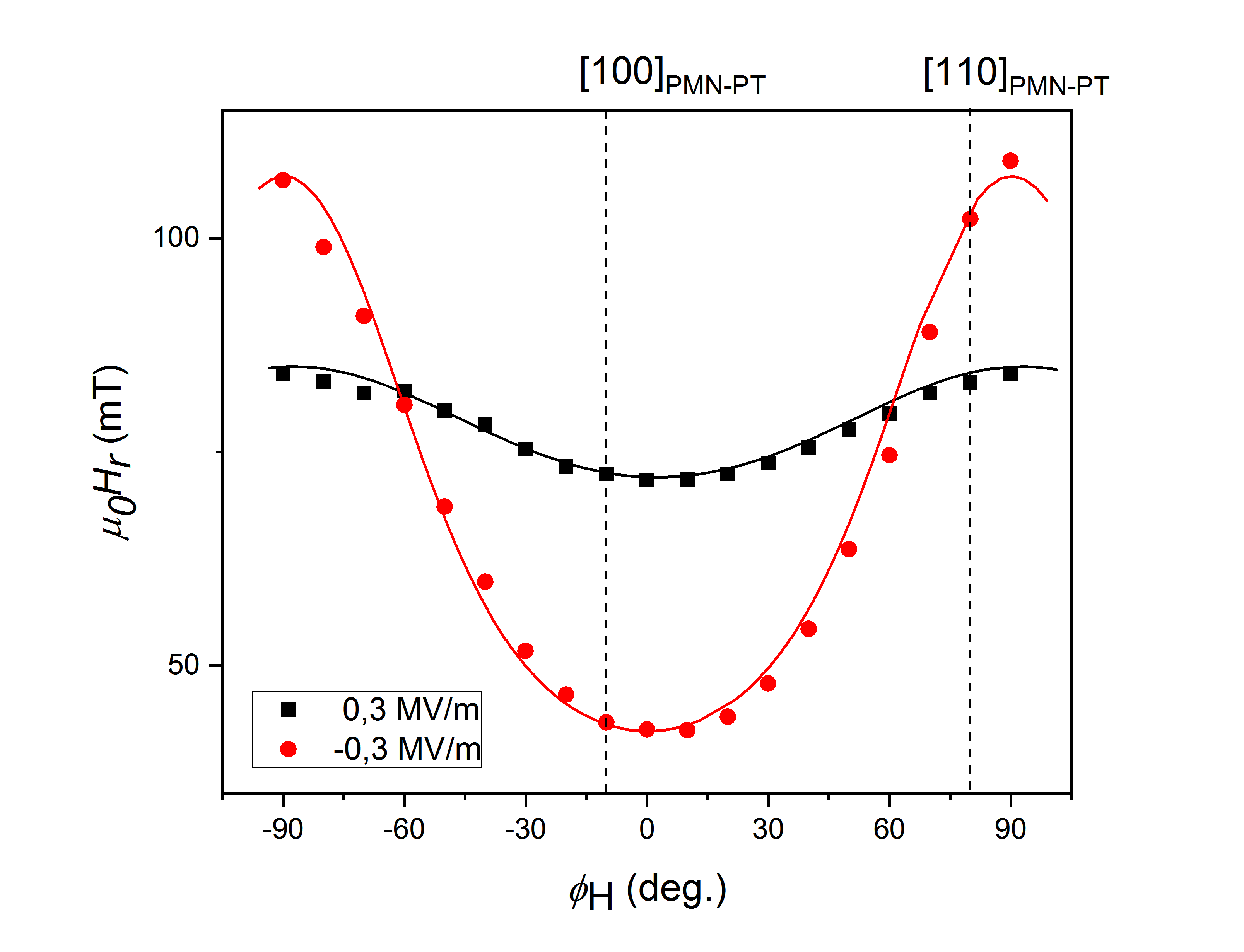}
\caption{Angular variation of the H$_r$ and fitting using the Smit-Beljers formalism and Eq.~\eqref{eq:Free_energy}.}
\label{fig:Hr_angular}
\end{figure}

The anisotropy constant values presented in Table~ \ref{tab:anisotropy_constants} are determined from fitting the experimental H$_r$($\theta,\varphi$) curves (Fig. \ref{fig:Hr_angular}) by solving the equation system composed by the Smit-Beljers formalism equation~\cite{G.A.Ramirez2021}:
\begin{equation}
\label{eq:Smit-Beljers}
 \left(\frac{\omega}{\gamma}\right)^2= \frac{1}{M_s^2 \textnormal{sin}^2 \theta} 
\left. \left[\frac{\partial^2  U}{\partial^2 \theta } \frac{\partial^2U}{\partial^2 \phi 
}-\left(\frac{\partial^2 U}{\partial \theta \partial \phi}\right)^2\right] 
\right \arrowvert_{\theta_{eq},\phi_{eq}},
\end{equation}

and Eq.~\eqref{eq:Free_energy} in a self-consistent scheme.


\begin{thebibliography}{<50>}

\bibitem{Ramesh2007} R. Ramesh and N. A. Spaldin, Nature Mater. {\bf6 },21 (2007).
\bibitem{Vaz2010} C.A.F. Vaz, F., J. Hoffman, C. H. Ahn, and R. Ramesh,  Adv. Matter. {\bf22 },2900-2918 (2010).
\bibitem{Bibes2008} M.Bibes and A. Barthelemy,  Nature Mater {\bf 7}, 425 (2008).
\bibitem{Hu2010} J.-M. Hu, Z.Li, and C. W. Nan, J. Appl.Phys., {\bf107}, 093912 (2010).
\bibitem{Liu2011}M.Liu, S. Li, O.Obi, J. Lou, S. Rand, and N.X. Sun, Appl. Phys. Lett {\bf98}, 222509 (2011).
\bibitem{Hu2010_2} J. -M Hu, Z. Li, Y. H. Lin, adn C. W. Nan, Phys. Status Solidi RRL, {\bf4}, 106 (2010).
\bibitem{Liu2009} M. Liu, O. Obi, J. Lou, Y. J. Chen, Z. H. Cai, S. Stoute, M. Espagnol, M. Lew, X, Situ, K. S. Zimer, V. G. Harris, and N. X. Sun, Adv. Func. Mater. {\bf19 }, 1826 (2009).
\bibitem{Shen2021}J.-X. Shen, H. Li, W.-H. Wang, S.-W. Wang and Y. Sun. Appl. Phys. Lett.{\bf119 } 192902 (2021).
\bibitem{Brandlmaier2011} A. Brandlmaier, S. Gerprägs, G. Woltersdorf, R. Gross, and S. T. B. Goennenwien, J. Appl. Phys.{\bf110}, 043913 (2011).
\bibitem{Eerenstein2007} W. Eerenstein, M. Wiora, J. L. Prieto, J. F. Scott, and N. D. Mathur,  Nature Mater{\bf6}, 348 (2007).
\bibitem{Thiele2007} C. Thiele, K. Dorr, O. Bilani, J. Del, and L. Shultz,  Phys. Rev. B. {\bf75}, 054408 (2007).
\bibitem{Heidler2015} J. Heidler, C. Piamoteze, R. V. Chopdekar, M. A. Uribe-Laverde, A. Alberca, M. Buzzi, A. Uldry, B. Delley, C. Bernhard, and F. Nolting, Phys. Rev. B {\bf 91} 024406 (2015). 
\bibitem{Yang2009} J. J. Yang, Y. G. Zhai, H. F. Tian, L. B. Luo, H. Y. Zhang, Y. J. He, and H. S. Luo, Appl. Phys. Lett.  {\bf 94}, 212504 (2009).
\bibitem{Wang2010-2011} J. Wang, J. -M. Hu, H. Wang, H. Jiang, Z. B. Wu, J. Ma, X. H. Wang, Y. H. Lin, and C. W. Nan, J. Appl. Phys. {\bf107}, 083901 (2010); J. Wang, H. Wang, H. Jiang, X. H. Wang, Y. H. Lin, and C. W. Nan J. Nanomater. 2010, 142750 (2010); J. Wang, J. Ma, Z.Li, Y. Shen; Y. H, Lin, and C. W. Nan, J. Appl. Phys. {\bf110}, 043919 (2011).
\bibitem{Wu2011} T. Wu, A. Bur, P. Zhao, K. P. Mohanchandra, K. Wong, K. L. Wang, C. S. Lynch, and G. P. Carman, Appl. Phys. Lett {\bf98 }, 012504 (2011).
\bibitem{Molegraaf2009} H. J. A. Molegraaf, J. Hoffman, C. A .F. Vaz, S. Gariglio, D. van der Mrel, C. H. Ahn, and J. M. Triscorne, Adv. Matter.  {\bf21 }, 3470 (2009).
\bibitem{Vaz2010-2} C. A. F. Vaz, J. Hoffman, Y. Segal, J. W. Reiner, R. D. Grober, Z. Zhang, C. H. Ahn, And F. J. Walker, Phys. Rev. Lett. {\bf104}, 127202 (2010).
\bibitem{Weisheit2007} M. Weisheit, Science {\bf104} 349 (2007).
\bibitem{Maruyama2009} T. Murayama, Y. Shiota, T. Nozaki, K. Ohta, N. Toda, M. Mizuguchi, A. A. Tulapurkar, T. Shinjo, M. Shiraishi, S. Mizukami, Y. Ando, and Y. Suzuki, Nat. Nanotech {\bf 4} 158 (2009).
\bibitem{Endo2010} M. Endo, S. Kanai, S. Ikeda, F. Matsukura, and H. Ohno, Appl. Phys. Lett. {\bf 96}, 117203 (2010).
\bibitem{Shu2012}L. Shu, Z. Li, J. Ma, Y. Gao, L. Gu, Y. Shen, Y. Lin, and C. W. Nan, Appl. Phys. Lett. {\bf 100}, 022405 (2012).
\bibitem{Nan2014}T. Nan, Z. Zhou, M. Liu, X. Yang et al, Sci. Rep. {\bf 4}, 3688 (2014).
\bibitem{Thiele2005} C. Thiele, K. Dörr, S. Fähler, and L. Schultz, Appl. Phys. Lett.  {\bf 87}  , 262502 (2005).
\bibitem{Heidler2016} J. Heidler, M. Fechner, R. V. Chopdekar, C. Piamonteze, J, Dreiser, C. A. Jenkins, A. Arenholz, S. Rusponi, H. Brune, and F. Nolting. , Phys. Rev. B {\bf 94}, 014401 (2016).
\bibitem{Clark2003}A. E. Clark, K. B. Hathway, M. W. Fogle, J. B. Restorff, T. A. Lograsso, V. M. Keppens, G. Petculescu, and R. A. Taylor, J. Appl. Phys. {\bf 93}, 8621 (2003).
\bibitem{Atulasimha2011} J.Atulasimha, A. B. Flatau, Smart Matter.Struct. {\bf 20} 043001 (2011). 
\bibitem{Meisenheimer2021}P. B. Meisenheimer, R. A. Steinhardt, S. H. Sung, L. D. Williams \textit{et al}, Nature Com. {\bf 12} 2757 (2021). 
\bibitem{Phuoc2017}N. N. Phuoc and C. K. Ong, J. Mater. Sci.:Mater. Electron. {\bf 28} 5628 (2017).
\bibitem{Zhang2018}Y. Zhang, C. Huang, M. Turghum, Z, Duan, F. Wang, and W. Shi, Appl. Phys. A {\bf 124} 289 (2018).
\bibitem{Bai2018}Y. Bai, N. Jiang, and S. Zhao, Nanoscale {\bf 10} 9816 (2018).
 \bibitem{Wu2011}T. Wu, A. Bur, K. Wong, P. Zhao, C. S. Lynch, P. Khalili Amiri, K. L. Wang, and G. P. Carman. Appl. Phys. Lett. {\bf 98} 262504 (2011).
\bibitem{Jimenez2019}M. J. Jimenez, G. Cabeza, J. E. Gomez, D. Velazquez Rodriguez, L. Leiva, J. Milano, and A. Butera, J. Mag. Mag. Mat. {\bf 501} 166361 (2020).
\bibitem{Jahjah2020}W. Jahjah, J.-Ph. Jay, Y. Le Grand, A. Fessant, A.R.E. Prinsloo, C.J. Sheppard, D.T. Dekadjevi, and D. Spenato, Phys. Rev. Applied {\bf 13} 034015 (2020).
\bibitem{Pradham2024}G. Pradhan, F. Celegato1, A. Magni, Marco Coisson, Gabriele Barrera,
Paola Rizzi, and Paola Tiberto,  J. Phys. Mater {\bf7} 015016 (2024).
\bibitem{Ramirez2021}G. A. Ramirez, A. E. Moya Riffo, J. E. Gomez, D. Goijman, L. M. Rodriguez, D. Frenegal, A. Butera, and J. Milano, Phys. Rev. B,{\bf 104}  064403 (2021). 
\bibitem{SIM_beamline}  U. Flechsig, F. Nolting, A. Fraile Rodríguez, J. Krempaský, C. Quitmann, T. Schmidt, S. Spielmann, and D.Zimoc,  AIP Conference Proceedings 1234, 319 (2010).
\bibitem{JStor2006}J. Störh and H. C. Siegman, \textit{Magnetism: From fundamentals to nanoscale dynamics}, edited by M. Cardona, P. Fulde, K. von Klizing, R. Merlin, H. J. Queisser, and H. Strömer, Solid-State Sciences  {\bf152} (Springer, New York, 2006). 
\bibitem{Duan2006}C.-G. Duan, S. S. Jaswal, E. Y. Tsymbal, Phys. Rev. Lett.{\bf97}, 047201 (2006).
\bibitem{crystal}J. Luo and S. Zhang, Crystals 4, 306 (2014)
\bibitem{Avula2018}S. R. V. Avula, J. Heidler, J. Dreiser, J. Vijayakumar, L. Howald, F. Nolting, C. Piamonteze. J. Appl. Phys. {\bf 123}, 064103 (2018).
\bibitem{Nataf2020} G. F. Nataf and E. K. H. Salje, Ferroelectrics  {\bf 569:1} 82 (2020).
\bibitem{Liu2010} M. Liu, O. Obi, Z. Cai, J. Lou, G. Yang, K. Ziemer, and N. X. Sun, J. Appl. Phys. {\bf 107} 073916 (2010). 
\bibitem{Chikazumi} S. Chikazumi, \textit{Physics of magnetism}, Chapter 8, Krieger Publishing Co, Florida, USA (1978).  
\bibitem{Yang2017} Y. T. Yang, J. Li, X. L. Peng, B. HOng, X. Q. Wang, H. L. Ge, D. H. Wang, and Y. W. Du, AIP Advances  {\bf7} 055833 (2017).  
\bibitem{110-textured} H. Ahmad, J. Atulasimha and S. Bandyopadhyay, Scientific Reports {\bf 5}, 18264 (2016). 
\bibitem{VESTA} Momma, Koichi and Izumi, Fujio, Journal of Applied Crystallography {\bf 41}, 653-658, (2008).
\end{thebibliography}

\begin{thebibliography}{<50>}
\bibitem{Regan}T. Regan, H. Ohldag, C. Stamm, F. Nolting, J. Lüning, J. Stöhr, and R. White, Physical Review B {\bf64}, 214422 (2001).
\bibitem{DFT} P.E. Blöchl; Phys. Rev. {\bf B 50}, 17953 (1994)
\bibitem{VASP1} G. Kresse and J. Furthmüller; Phys. Rev. {\bf B 54}, 11169 (1996)
\bibitem{VASP2} G. Kresse and D. Joubert; Phys. Rev. {\bf B 59}, 1758 (1999)
\bibitem{PBE} J. Perdew, S. Burke and M. Ernzerhof; Phys. Rev. Lett. {\bf 77} 3865 (1996)
\bibitem {sqs-1} A. Zunger, S.H. Wei, L.G. Ferreira and J.E. Bernard; Phys. Rev. Lett. {\bf 65}, 353 (1990)
\bibitem{sqs-2} A. van de Walle, P. Tiwary, M. de Jong, D.L. Olmsted, M. Asta, A. Dick, D. Shin, Y. Wang, L.-Q. Chen and Z.-K. Liu; CALPHAD: Computer Coupling of Phase Diagrams and Thermochemistry {\bf 42}, 13 (2013)
\bibitem{ForceTheorem} G. H. O. Daalderop, P. J. Kelly, and M. F. H. Schuurmans; Phys. Rev. {\bf B 41} , 1919, 1990. 
\bibitem{G.A.Ramirez2021}G.A. Ramírez, A. Moya-Riffo, D. Goijman, J.E. Gómez, F. Malamud, L.M. Rodríguez, D. Fregenal, A. Butera, J. Milano; J. Magn. Magn. Mater..{\bf535}, 16047, 2021.
\end{thebibliography}
\end{document}